\documentstyle[12pt,titlepage]{article}
\def\singlespace {\smallskipamount=3.75pt plus1pt minus1pt
                  \medskipamount=7.5pt plus2pt minus2pt
                  \bigskipamount=15pt plus4pt minus4pt
                  \normalbaselineskip=15pt plus0pt minus0pt
                  \normallineskip=1pt
                  \normallineskiplimit=0pt
                  \jot=3.75pt
                  {\def\smallskip {\vskip\smallskipamount}}
                  {\def\medskip   {\vskip\medskipamount}}
                  {\def\bigskip   {\vskip\bigskipamount}}
                  {\setbox\strutbox=\hbox{\vrule
                    height10.5pt depth4.5pt width 0pt}}
                  \parskip 7.5pt
                  \normalbaselines}
\def\middlespace {\smallskipamount=5.625pt plus1.5pt minus1.5pt
                  \medskipamount=11.25pt plus3pt minus3pt
                  \bigskipamount=22.5pt plus6pt minus6pt
                  \normalbaselineskip=22.5pt plus0pt minus0pt
                  \normallineskip=1pt
                  \normallineskiplimit=0pt
                  \jot=5.625pt
                  {\def\smallskip {\vskip\smallskipamount}}
                  {\def\medskip   {\vskip\medskipamount}}
                  {\def\bigskip   {\vskip\bigskipamount}}
                  {\setbox\strutbox=\hbox{\vrule
                    height15.75pt depth6.75pt width 0pt}}
                  \parskip 11.25pt
                  \normalbaselines}
\def\doublespace {\smallskipamount=7.5pt plus2pt minus2pt
                  \medskipamount=15pt plus4pt minus4pt
                  \bigskipamount=30pt plus8pt minus8pt
                  \normalbaselineskip=30pt plus0pt minus0pt
                  \normallineskip=2pt
                  \normallineskiplimit=0pt
                  \jot=7.5pt
                  {\def\smallskip {\vskip\smallskipamount}}
                  {\def\medskip   {\vskip\medskipamount}}
                  {\def\bigskip   {\vskip\bigskipamount}}
                  {\setbox\strutbox=\hbox{\vrule
                    height21.0pt depth9.0pt width 0pt}}
                  \parskip 15.0pt
                  \normalbaselines}
\begin{document}
\title{Baryogenesis via lepton number violating scalar interactions}
\author{Patrick J. O'Donnell \\
Physics Department \\
University of Toronto \\
Toronto, Ontario M5S 1A7, Canada\\
\\ and \\ \\
Utpal Sarkar \\
Theory Group \\
Physical Research Laboratory \\
Ahmedabad - 380 009, India}

\date{UTPT-93-15}

\maketitle

\begin{abstract}
\middlespace

We  study   baryogenesis   through  lepton  number  violation  in
left-right  symmetric  models.  In these models the lepton number
and $CP$ violating  interactions of the triplet higgs scalars can
give rise to  lepton  number  asymmetry  through  non-equilibrium
decays  of the  $SU(2)_L$  triplet  higgs  and the  right  handed
neutrinos.  This in turn generates  baryon  asymmetry  during the
electroweak anomalous processes.

\end{abstract}
\newpage
\middlespace

Cosmological  baryon excess can be generated from
the initial condition $B=0$ if there is out-of-equilibrium baryon
number  violation along with CP violation  \cite{kolb}.  However,
the sphaleron induced, anomalous  electroweak,  process will wash
out all the $(B - L)$  conserving  baryon  asymmetry  \cite{anom}
which is  generated  at the grand  unified  scale.  The  ratio $$
\frac{n_B}{n_\gamma} = (4 - 7) \times 10^{-10} $$ required by the
standard  big-bang  model, can then be  explained  if there  were
primordial  $(B - L)$  asymmetry.  Since the  lowest  dimensional
operators for the proton decay  conserve  $(B-L)$, it is unlikely
that  grand  unified   theories  can  generate   enough   $(B-L)$
asymmetry.  Two interesting  possibilities  then remain open.  In
the first one the observed  asymmetry is generated just after the
electroweak  phase  transition  \cite{brahm}.  In this  case  the
electroweak  phase  transition  has to be strongly first order so
that after the  transition  the anomalous  process is already too
weak to erase the generated asymmetry.  The second possibility is
the one where $(B-L)$ asymmetry is generated during the anomalous
electroweak  process.  If  there  are  lepton  number  violating,
out-of-equilibrium,  interactions  which  violate CP, then during
the epoch when the $(B+L)$ violating  anomalous  processes are in
equilibrium  in the universe, this lepton  asymmetry can generate
enough $(B - L)$ asymmetry.  Some time back Fukugita and Yanagida
\cite{fy,luty,model}  proposed this possibility,  where the right
handed  neutrinos  decay into light  leptons and  antileptons  in
different proportions when CP is violated \cite{us}.  This lepton
asymmetry in turn generates baryon asymmetry.  They considered an
extension  of the standard  model which  contains  singlet  right
handed  neutrinos  and the lepton number is broken  explicitly by
the Majorana mass terms of the right handed neutrinos.

In this article we discuss the possibility of baryon asymmetry in
models  where  $(B - L)$  is  broken  spontaneously  rather  than
putting in an explicit $(B - L)$ violating  interaction.  This is
natural in the  left-right  symmetric  extension  of the standard
model  \cite{lrm}.  The  spontaneous  breaking  of $(B - L)$ give
rise to lepton number violating interactions of the triplet higgs
scalars.  This can then  generate  lepton  asymmetry  if there is
also $CP$  violation  through  decays of the higgs  and the heavy
neutrinos.  This  generates  enough baryon  asymmetry  during the
electroweak anomalous process.  We use the conventional choice of
the higgs  scalars for the breaking of the  left-right  symmetry,
which also gives Majorana mass to the neutrinos.  We also discuss
an   alternative,   where   the   left-right   parity  is  broken
spontaneously \cite{par}.

In  the  left-right  symmetric  model  \cite{lrm},  the  standard
electroweak   theory  $G_{std}  \equiv  SU(3)_c  \otimes  SU(2)_L
\otimes  U(1)_Y$  emerges at low  energy as a result of  symmetry
breaking of a larger group $G_{LR} \equiv SU(3)_c \otimes SU(2)_L
\otimes  SU(2)_R  \otimes  U(1)_{B-L}$.  The  vacuum  expectation
value  ({\it  vev})  of the  right  handed  triplet  higgs  field
$\Delta_R  $  (1,1,3,-2),  breaks  the gauge  group  $G_{LR}  \to
G_{std}$.  Left-right  symmetry  implies the existence of another
higgs field  $\Delta_L $ which  transforms  as  (1,3,1,-2)  under
$G_{LR}$.  A higgs  doublet  field  $\phi$  (1,2,2,0)  breaks the
electroweak  symmetry  and  gives  masses  to the  fermions.  The
Yukawa  couplings are
\begin{equation}
{\cal  L}_{Yuk} = f_{ij}  \overline{\psi_{iL}}  \psi_{jR}  \phi
+ f_{Lij}  \overline{{\psi_{iL}}^c}  \psi_{jL}  \Delta_L^\dagger
+ f_{Rij}  \overline{{\psi_{iR}}^c}  \psi_{jR} \Delta_R^\dagger.
\label{Yuk}
\end{equation}
The scalar  interactions which are of
importance  for  the  generation  of  lepton  number  excess  are
\begin{equation}
{\cal L}_{int} = g  (\Delta_L^\dagger  \Delta_R \phi \phi
+ \Delta_L  \Delta_R^\dagger \phi \phi ) + h.c. \label{scalar}
\end{equation}
The  minimization  of  the  complete  scalar  potential  gives  a
relation  between the {\it vev}\,s of these  fields.  Since right
handed  gauge  bosons  have  not  been   observed,  we  are  only
interested  in $v_L \neq v_R$,  where  $v_{L,R}$  and $v$ are the
{\it   vev}\,s   of  the   fields   $\Delta_{L,R}$   and   $\phi$
respectively.  If the  left-right  parity  ($D-$  parity)  is not
broken, then the masses of the fields  $\Delta_L $ and $\Delta_R$
remains  same even after the  breaking of  $G_{LR}$,  {\it i.e.,}
$m_{\Delta_L}  =  m_{\Delta_R}  =  m_{\Delta}  \approx v_R$.  The
{\it
vev}\,s of the fields  $\Delta_{L,R}$  and $\phi$ are  related by
$v_L \approx {v^2 / v_R}$.  This allows a see-saw  suppression of
the {\it vev} of $\Delta_L$ compared to $\Delta_R$.

In these models $(B - L)$ is a local  symmetry.  The  breaking of
the group  $G_{LR}$ gives  spontaneous  breaking of the $(B - L)$
symmetry.  Thus {\it vev}\,s of the higgs scalars  $\Delta_{L,R}$
give rise to baryon and lepton number  violating  interactions by
two units, which can allow  Majorana  mass of the  neutrinos  and
also  neutron-antineutron  oscillations.  In the model considered
by Fukugita  and Yanagida  \cite{fy},  there was a Majorana  mass
term for the neutrinos  which violated  explicitly  lepton number
conservation.  Such lepton number  violating  interactions  gives
rise to baryogenesis  via  leptogenesis  through the decay of the
heavy  right  handed  neutrinos.  In  the  left-right   symmetric
theories there exists other processes which arise from the lepton
number  violating  interactions  of the scalar  fields.  They can
allow lepton asymmetry through the  non-equilibrium  decay of the
left-handed  higgs triplet and the right handed  neutrinos, which
then generates baryon asymmetry through sphaleron processes.

Following Eq. (\ref{Yuk}), $\Delta_{L,R}$  can decay
into two neutrinos, $\Delta_{L,R}^\dagger$ into two antineutrinos
\begin{eqnarray}
  \Delta_{L,R}  &\to& \nu_{L,R} + \nu_{L,R} \label{eqna} \\
 \Delta_{L,R}^\dagger  &\to& \nu_{L,R}^c + \nu_{L,R}^c \label{eqnb}
\end{eqnarray}
Equation (2) gives lepton number  violating scalar  interactions of
$\Delta_{L}$ and $\Delta_{R}$,  which is generated by the $(B-L)$
violating  {\it  vev}\,s of  $\Delta_{R,L}$.  Together  with
(\ref{eqna})  and   (\ref{eqnb}),   these  lepton  number   violating
interactions then give
$$ \Delta_{L,R} \to \phi + \phi
$$ $$ \phantom{\Delta_{L,R}}  \to \phi^\dagger + \phi^\dagger .$$
To generate excess lepton number however, we need to violate $CP$
and  have  an  imaginary   one-loop   radiative   correction.  In
principle, there can be a, rephasing  invariant, complex phase in
the  Yukawa   couplings,  which  can  give  $CP$  violation.  The
interference  of the tree level  diagram and the one loop diagram
of Fig.1 can then give rise to an asymmetry in the decay modes of
(\ref{eqna})  and  (\ref{eqnb})  for  the  left-handed   triplets
$\Delta_L$.  For the right-handed triplet $\Delta_R$ the loop
integral is real since the left-handed  neutrinos, which enter in
the  loop,  are  light.  This  implies  that  the  decay  of  the
right-handed  triplet higgs cannot generate any lepton asymmetry.
The  magnitude  of the  asymmetry  generated  by the  left-handed
triplet $\Delta_L$ decays is given by,
\begin{equation}
\epsilon_\Delta  \approx  \frac{1}{4  \pi  |f_{L,R}|^2}
{\rm  Im}  [g^* f_{[L,R]ij}^*  f_{ik}  f_{jk}]
\frac{g^* }{f_{[R,L]kk}} \label{eps}
\end{equation}
In general, the quantity $[g^* f_{[L,R]ij}^*  f_{ik} f_{jk}]$ can
contain a, rephasing  invariant,  $CP$ violating phase and so can
be complex \cite{us}.  This generates the lepton number excess in
the decays of the $SU(2)_L$  higgs  triplets  $\Delta_{L}$.  Note
that in the loop integral the {\it vev}  $(v_{R})$ of the triplet
field $\Delta_{R}$  enters in the coupling of the $\Delta_{L}$ to
$\phi$ and cancels with that in the propagator of the  neutrinos.
Hence the decays of  $\Delta_L$  are not  suppressed  by any {\it
vev}\,s.

Lepton  number  asymmetry  is also  generated  by the heavy right
handed   neutrino   decays.   The   {\it   vev}   of   $\Delta_R$
spontaneously   gives  a  Majorana   mass  to  the  right  handed
neutrinos.  This in turn  allows  the  decay  of  $\nu_R$  into a
lepton {\it and} an antilepton,
\begin{eqnarray}
  \nu_{iR}  &\to& l_{jL} + \bar{\phi} \label{eqn1a} \\
   &\to&  {l_{jL}}^c + {\phi} .\label{eqn1b}
\end{eqnarray}
In the case of right-handed neutrinos there are two types of loop
diagrams  which  can  interfere  with the tree  level  decays  of
(\ref{eqn1a})  and  (\ref{eqn1b})  which are shown in Fig.2.  The
interference  of the tree level  diagram and the one loop diagram
of Fig.2(a)  generates lepton asymmetry of a magnitude similar to
that  of  the  triplet  higgs  decay.  There  is  one  additional
contribution   to  the   lepton   asymmetry   generated   by  the
interference  of the tree  level  diagram  and that of  Fig.2(b).
This second  contribution  is similar to the ones  studied in the
earlier  papers on  leptogenesis,  where lepton  number is broken
explicitly   through  the   Majorana   mass  term  of  the  right
handed-neutrinos  \cite{fy,luty,us,model}.  The magnitude of this
lepton number excess is given by,
\begin{equation}
\epsilon_\nu  \approx  \frac{1}{4  \pi  |f_{ik}|^2}  {\rm  Im}  [f_{ik}
f_{il}  f_{jk}^*  f_{jl}^*]  \frac{f_{Rii}}{f_{Rkk}} .\label{eps1}
\end{equation}

For the generation of the lepton number asymmetry we
require another ingredient, namely, the decay rates should satisfy
the out-of-equilibrium limit. The process
(say, $X$-decay) which generates the lepton asymmetry should  satisfy
\begin{equation}
 \mbox{} \hskip 1.8in \Gamma_X \leq 1.7 \sqrt{g} \displaystyle
 \frac{T^2}{M_{Pl}}
 \hskip 1in {\rm at} \;\;\; T=M_X  \label{const}
\end{equation}
and  should  decouple  after  the  other  particles  whose  decay
violates  lepton  number have already  decayed  away and no other
lepton number violating  processes are in equilibrium.  Otherwise
after the  $X$-decay  has  generated  the  asymmetry  and already
decayed, the other processes in equilibrium  will again erase the
asymmetry  generated  by the  $X$-decay.

We now have to consider only the $\Delta_L, \; \Delta_R  \;
{\rm and} \;\;\; \nu_{1R}$ decay processes (we assume $\nu_{1R}$
to be the  lightest  of the right  handed  neutrinos).  The decay
widths for $\Delta_{L,R}$ and $\nu_{1R}$ are,
\begin{equation}
\Gamma_{\Delta_{L,R}} =  \displaystyle \frac{| f_{[L,R]ij} |^2}{16 \pi}
 M_\Delta \hskip .4in {\rm and  } \hskip .4in
\Gamma_{\nu_{1R}} =  \displaystyle \frac{| f_{1j} |^2}{16 \pi}
 M_{\nu}
\end{equation}
where $M_\nu$ is the mass of $\nu_{1R}$.  Since the masses of all
these  fields are of the same order of  magnitude  $\sim v_R$, we
have to consider all of them for the  understanding of the lepton
number  generation.

If  $\Delta_R$  is  heavier  than the other  particles,  then the
lightest of  $\Delta_{L}$  or $\nu_R$ (with mass $m$, say) should
satisfy the out-of-equilibrium  condition.  Then, at $T = m$, all
other fields have decayed  away since the inverse  reactions  are
not allowed by phase space.  The lepton  number is generated by the
lightest   of   $\Delta_{L}$   or   $\nu_R$.  However,   if  both
$\Delta_{L}$  and $\nu_R$  satisfy  (\ref{const})  then the total
lepton  number  generated  is the  sum of the  contribution  from
$\Delta_{L}$  and  $\nu_R$.

On the other hand if  $\Delta_R$  is lighter, then in addition to
the   $\Delta_{L}$  or  $\nu_R$   (whichever  is  the  lightest),
$\Delta_R$ should also satisfy the out-of-equilibrium  condition.
Otherwise it will wash out the lepton number excess  generated by
the other particles.  Although $\Delta_R$ can not generate lepton
asymmetry since the loop integral is real, its equilibrium  decay
can  make the  number  of  leptons  to be same as the  number  of
antileptons.

If  the  Yukawa   coupling  for  the  first
generation of neutrinos is of the order of $\sim  10^{-5}$,  then
the  out-of-equilibrium  condition requires $M > 10^6$ GeV.  This
constraint  is  satisfied  by most  of the  left-right  symmetric
models.  However, if left-right  symmetry is observed at around a
few TeV, then the baryon  asymmetry  has been  generated  through
some other mechanism or else the Yukawa couplings for the triplet
higgs should be $< 10^{-7}$.  If the out-of-equilibrium condition
is  satisfied,  then  the  next  question  will  be  whether  the
$\epsilon$\,s  can be of the right order of magnitude to generate
the observed baryon asymmetry.  However, given the uncertainty in
the  Yukawa  and  the  quartic  scalar  couplings,  it is  always
possible  to find a  suitable  choice  of  parameters  which  can
produce the right amount of baryon asymmetry.

There is another  variation of the left-right  symmetric model in
which the left-right  $D$-parity is broken  spontaneously.  Under
$D$-parity  the scalar  and the  fermionic  fields  transform  as
$\Delta_{L,R} \to \Delta_{R,L}$  and $\psi_{L,R} \to \psi_{R,L}$,
while $\phi$ stays the same.  This  $D$-symmetry can be broken by
the  {\it  vev} of the  singlet  field  $\eta$  (1,1,1,0),  which
transforms  under $D$ as $\eta \to - \eta$.  In the  presence  of
the field $\eta$ the lagrangian will now contain new terms,
$$
{\cal L}_{\eta \Delta} = M_{\eta} \eta (\Delta_L^\dagger \Delta_L
  - \Delta_R^\dagger \Delta_R) + \lambda_{\eta}  \eta^2
  (\Delta_L^\dagger  \Delta_L + \Delta_R^\dagger \Delta_R)
$$
which can then allow a different scenario for the {\it vev}\,s
and masses.  One can find  a solution of the form
\cite{par}
$$
v_\eta = \langle \eta \rangle \gg v_R \gg v_L \;\;\;\;\;\;
{\rm and}
\;\;\;\;\;\; v_L \approx {v^2 \over \langle \eta \rangle}
$$
$$
M_\eta \approx  M_{\Delta_L}  \approx  v_\eta \gg  M_{\Delta_R} =
M_\Delta \approx v_R $$ Once this $D$-parity is broken, the gauge
coupling  constants $g_L$ and $g_R$ for the groups  $SU(2)_L$ and
$SU(2)_R$,  respectively,  will evolve in a different  way (since
$M_{\Delta_L}  \gg  M_{\Delta_R}$)  and hence at low energy, even
before $G_{LR}$ is broken, $g_L \neq g_R$.  Similarly, the Yukawa
couplings $f_L$ and $f_R$ can also differ.

The  $\Delta_L$   mass  is  very  high.  If
$\Delta_R$ is now in  equilibrium,  then only  out-of-equilibrium
decays of $\nu_R$ can generate  the  asymmetry.  Otherwise  if
$\Delta_{L,R}$ and    $\nu_R$   satisfy  the  constraint
(\ref{const}) then both $\Delta_L$ and $\nu_R$ will contribute to
the generation of lepton asymmetry.

In  all  the  scenarios   considered   here,  the  higgs  scalars
$\Delta_L$ and $\Delta_R$ and their  interactions introduce
new processes at low energies such as $$\nu_L + \nu_L \to \nu_L^c
+ \nu_L^c  \hskip .5in {\rm and } \hskip .5in \nu_L +  \bar{\phi}
\to  \nu_L^c + \phi .  $$ The rates for these  processes  are
suppressed  by   $M_\Delta^{-4}  $ and
$M_\nu^{-2}$.  These  again  wash out the  baryon
asymmetry  unless their rates are smaller than the expansion rate
of the  universe  at the time of  electroweak  phase  transition.
These give   further   bounds  on  the  mass  scales
\cite{fyb,camp};
$$ M_{\Delta_L} > f_{L} 10^{6} \left(\frac{T_c}{100 \;{\rm GeV}}
\right)^{3/4}
{\rm GeV} $$ $$M_{\nu} > f^2 10^{9} \left(\frac{T_c}{100
\;{\rm GeV}}\right)^{1/2} {\rm GeV} ,$$
which are much weaker than the
bounds on the mass  scales
from the non-equilibrium condition (\ref{const}).

In summary, we discussed the possibility of baryogenesis  through
lepton number  violation in  left-right  symmetric  theories.  In
this case the lepton number  violating  decays of the left-handed
triplet  higgs or the right  handed  neutrinos  can generate  the
lepton asymmetry during the electroweak  anomalous  process which
can then generate  baryon  asymmetry.  On the other hand if these
lepton number  violating  processes are in equilibrium  then they
can wash out any primordial $(B - L)$ asymmetry.

{\bf  Acknowledgment}  We would like to thank the NSERC of Canada
for an International  Scientific  Exchange Award.

\newpage
\middlespace

\vskip 1in

\section*{Figure Captions}

\vskip .3in
\begin{itemize}
\item[Figure 1] Tree level and one loop diagrams for the decay of the
left-handed
triplet higgs scalars which generate lepton asymmetry \\
\vskip .2in
\item[Figure 2] One loop diagrams for the decay of the right-handed neutrinos
which generate lepton asymmetry
\end{itemize}

\end{document}